\documentclass[hidelinks,twocolumn, prl, superscriptaddress,a4paper]{revtex4}
\usepackage[colorlinks=true,allcolors=blue]{hyperref}
\usepackage{graphicx}% Include figure files
\usepackage{dcolumn}% Align table columns on decimal point
\usepackage{orcidlink}
\usepackage{csquotes}
\usepackage{float}
\usepackage{amsmath}
\usepackage{bm}% bold math
\usepackage{xcolor}
\usepackage{comment}

\begin{document}

\preprint{APS123-QED}

%\preprint{}


% \affiliation can be followed by \email, \homepage, \thanks as well..
\author{Maitri Ganguli\,\orcidlink{0009-0009-4701-2459}}
%\thanks{}
\affiliation{Department of Physics, Indian Institute of Science, Bangalore}


\author{Aneek Jana\,\orcidlink{0009-0001-1097-4250}}

\affiliation{Undergraduate Program, Indian Institute of Science, Bangalore}

\author{Awadhesh Narayan\,\orcidlink{0000-0003-0127-7047}}

\affiliation{Solid State and Structural Chemistry Unit, Indian Institute of Science, Bangalore}

\title{Floquet engineering of topological semimetals with bicircularly polarized light}



\date{\today}

\begin{abstract}
We study the effect of illumination of bicircularly polarized light (BCL) on topological semimetals, namely multifold fermion and line-node semimetals. We demonstrate, by means of both low-energy and lattice models, that the band-structures of both multifold fermions and line-node semimetal exhibit rich features with BCL illumination. For multifold fermions, we show that if the relative amplitude, $r$, between the two frequency components of the BCL is related to the relative frequency, $\eta$, in such a way that $r^2 = \eta$, then the band touching points remain unchanged without any shift or band gap opening for any choice of the material-dependent parameters. For other choices of the BCL parameters, we find both gapped phases as well as gapless phases with shifted band touching points. Line-node semimetals also exhibit diverse phases, ranging from protected line-nodes under illumination to point nodes, by tuning the BCL parameters $r$ and $\eta$. Overall, our analytical and numerical results establish BCL as a versatile tool for obtaining tunable topological phases.
\end{abstract}

%\keywords{}

\maketitle

\section{Introduction}

In recent years, Floquet engineering has come to fore in various domains of physics. The core idea of this technique is to control and manipulate systems by periodically driving them~\cite{pdrive1,pdrive2,pdrive3,pdrive4}. This facilitates revealing new dynamical regimes and altering the system's effective properties in ways that are not possible with static perturbations alone~\cite{rudner2019floquettopologicalinsulatorsband}. For instance, Floquet engineering has profound implications on the material properties~\cite{FQM1,FQM2,FQM3}. Furthermore, in ultracold atomic systems~\cite{ucold1,uc1,uc2} and optical lattices, periodic modulations have been used to simulate complex quantum phenomena and investigate exotic phases of matter~\cite{qpm1,qpm2,qpm3,qpm4,qpm5}.

On the other hand, topological semimetals are an exciting class of materials that have garnered significant attention in condensed matter physics and materials science~\cite{Gao_2019,Burkov_2016,Yan_2017,RevModPhys.93.025002}. Such systems have either discrete points or lines in momentum space where the valence band and conduction band touch, forming gapless electronic excitations. Unlike conventional semimetals or metals, the electronic states in topological semimetals are protected by topological invariants, making their properties robust against certain types of perturbations~\cite{RevModPhys.82.3045}. An upcoming subclass show higher number (more than 2) of bands touching, commonly known as multifold semimetals~\cite{Bradlyn_2016,PhysRevLett.116.186402,mf2,mf3}. The presence of such degeneracies has already been confirmed in experiments~\cite{PhysRevLett.119.206401,PhysRevLett.122.076402,Rao_2019}. As multifold fermions show enhanced linear response compared to Weyl fermions~\cite{S_nchez_Mart_nez_2019}, these kinds of semimetallic systems are of great interest. On the other hand, there are certain gapless systems that exhibit band touching along lines and result in interesting loop-like structures of the Fermi surfaces -- these are termed line node semimetals~\cite{PhysRevB.84.235126}. Such symmetry-protected topological semimetals have been reported in several experimental and theoretical studies~\cite{PhysRevB.92.045108,PhysRevLett.115.026403,PhysRevLett.115.036806}.

Floquet engineering has emerged as a promising method for engineering the band structures of topological semimetals~\cite{Wang_2014,PhysRevB.91.205445,PhysRevLett.117.087402,ANLineNodeCL,PhysRevB.94.121106,H_bener_2017} through time-periodic external fields, such as applying laser light. In this regard, the application of circularly polarized light (CL) or elliptically polarized light is most common. For example, it is possible to induce topological phase transitions by breaking time-reversal symmetry in Dirac or Weyl semimetals using CL. These ideas have led to various theoretical predictions and experimental observations such as the anomalous Hall effect~\cite{PhysRevLett.116.026805} and appearance of exotic surface states termed Floquet-Fermi arcs~\cite{PhysRevB.110.L121118}. Floquet engineering allows for dynamic control over the band structure, through which it has been possible to realize novel electronic phases that may not exist in equilibrium conditions.

While the interplay of CL with topology has been intensively studied, the effects of BCL on topological properties are relatively unexplored. A handful of early studies involving the interaction of BCL with quantum matter have very recently appeared~\cite{PhysRevA.102.063315,DW2,PhysRevResearch.4.013056,PhysRevA.107.043309,bae2024topologicalfloquetengineeringthreeband,RAJPOOT2024129241}. These include application of BCL to Dirac and Weyl semimetals~\cite{DW2,DW1}, tunable breaking of symmetries~\cite{lattice1}, and controllable photocurrents~\cite{photo1}.

\begin{figure}[t]
    \centering
    \includegraphics[width=0.48\linewidth]{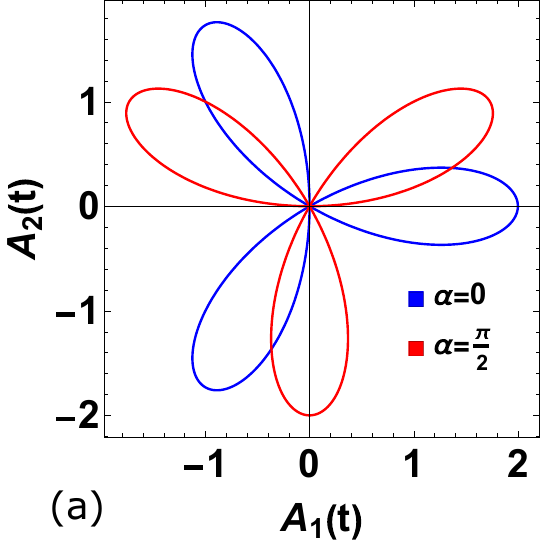}
    \includegraphics[width=0.48\linewidth]{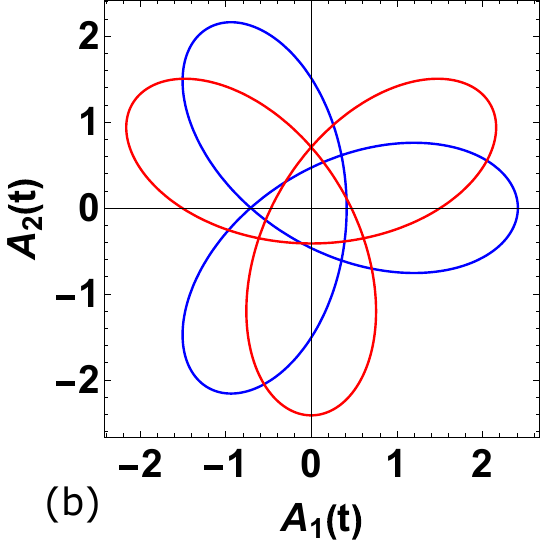}  
    \caption{\textbf{BCL vector potential for different amplitude ratios}. The vector potential for BCL with \(\eta=2\) with relative amplitudes (a) \(r=1\) and (b) \(r=\sqrt{2}\). Different colors represent two different values of phase difference $\alpha$. We have set \(\omega=1\) and \(\mathcal A_0 = 1\). Changing \(r\) and \(\alpha\) can tune the shape and orientation of the applied BCL.}
    \label{fig:BCL rosette}
\end{figure}

In this work, we explore the effect of illumination of BCL in multifold fermions and line node semimetals. Using the frequency-ratio \(\eta\), the amplitude ratio \(r\), and the relative phase \(\alpha\) between the two CLs comprising the BCL, as tuning parameters, we analytically show that the topological band structures can be manipulated in interesting ways resulting in high tunability. For multifold fermions, the presence and absence of a band touching point (BTP) as well as the position of the BTP can be controlled in more versatile ways than applying ordinary CL. In particular, for a specific choice of relative amplitude, $r$, and relative frequency, $\eta$, (\(r^2=\eta\)), the BTP remains intact at its original position in the parent model. On the other hand, for line-node semi-metals, the transition to point nodes under illumination can be controlled by tuning the above-mentioned BCL parameters. For certain choice of the parameters (\(r^2=\eta=2 \text{ with } \alpha=0\) or \(r^2=\eta>2\)), the line-node band structure can be kept intact even under non-zero overall illumination amplitude. Furthermore, we find that when the frequency ratio \(\eta=2\) and relative phase \(\alpha\neq 0,\pi\), then the resulting band structure will depend on the quadratic part of the line-node dispersion. To be specific, if it is isotropic, we can obtain a point-node or a line-node depending on the value of relative amplitude \(r\), and if it is anisotropic, then a band gap will open up. We have complemented our analytical analysis by numerical calculations on lattice models of both multifold fermions and line-node semimetals. It would be interesting to verify our predictions in experiments. 

\section{Floquet engineering with BCL}

Let us begin by examining how we can obtain BCL using circularly polarized light, depending on the relative phase differences between the circularly polarized light beams, their relative frequencies, and amplitude ratios. Further we discuss briefly the Floquet-Magnus expansion for BCL, which we use in the later sections.

\subsection{A. Vector potential of BCL}

BCL consists of two CL beams with integer frequency ratio \(\eta\) and opposite polarizations. In previous work~\cite{DW2}, two CL beams with equal amplitude were superposed to obtain the BCL. In our work, we introduce additionally the amplitude ratio of the superposed CL beams, $r$, which allows further possibilities for tuning the properties of different systems, as we shall see. Thus, we consider the following vector potential with a time-period \(T=2\pi/\omega\),

\begin{equation}
    \mathbf A (t) = \sqrt{2} \mathcal A_0 \text{ Re}\left(r e^{-i(\eta\omega t-\alpha)} \hat \varepsilon_R + e^{-i\omega t} \hat \varepsilon_L\right),
\end{equation}

where \(\mathcal A_0\) is an overall amplitude and \(\alpha\) is a relative phase between the CLs. Here \(\hat \varepsilon_R\) and \(\hat \varepsilon_L\) are two opposite circular polarizations in the plane of two orthonormal unit vectors \(\hat \varepsilon_1\) and \(\hat \varepsilon_2\),

\begin{equation}
    \hat \varepsilon_R = \frac{1}{\sqrt{2}}(\hat \varepsilon_1-i\hat \varepsilon_2),\,\,\hat \varepsilon_L = \frac{1}{\sqrt{2}}(\hat \varepsilon_1+i\hat \varepsilon_2).
\end{equation}

In component form, the vector potential is,

\begin{equation}
    \begin{aligned}
        &A_1 (t) = \mathcal A_0 (r \cos (\eta\omega t - \alpha) + \cos (\omega t)),\\
        &A_2 (t) = \mathcal A_0 (-r \sin (\eta\omega t - \alpha) + \sin (\omega t)).
    \end{aligned}
\end{equation}

Different choices of \(r\) yield different rosette patterns of the tip of vector potential, and the effect of \(\alpha\) is an overall angular shift, as shown in Fig.~\ref{fig:BCL rosette}. Illumination of this BCL amounts to a Peierls substitution in the momentum space Hamiltonian, that is \(H(\mathbf k) \rightarrow H (\mathbf k + e\mathbf A(t))\). This results in a time-periodic Hamiltonian \(H(t)\) with time period \(T\), \(H(t+T) = H(t)\). Note that for simplicity we will set the electronic charge, $e=1$.

\subsection{B. Floquet-Magnus expansion}

Let us briefly recall the Floquet-Magnus expansion which we will employ in our analysis. In the Floquet-Magnus approximation, the time-dependent Hamiltonian is replaced by an effective time-independent Hamiltonian in a high frequency expansion in powers of $1/\omega$. We begin by calculating the $m$ Fourier modes of the time-dependent periodic Hamiltonian,

\begin{equation}
    H_m = \frac{1}{T} \int_0^T dt \, H(t) \, e^{-im\omega t}.
\end{equation}

In terms of these Fourier modes, the leading-order effective Floquet Hamiltonian becomes (with \(\hbar=1\))~\cite{Oka_2019},

\begin{equation}
\begin{aligned}
    H_{\text{Floq}} &= H_0 - \sum_{n=1}^\infty \frac{1}{n\omega} \left([H_{-n},H_n]\right)\\
    &+\sum_{n=1}^\infty \frac{1}{2n\omega} \left([H_0,H_n]-[H_0,H_{-n}]\right) +\mathcal{O}\left(\frac{1}{\omega^2}\right).
\end{aligned}
\end{equation}

The above Floquet-Magnus expansion is valid when \(H_m/\omega\) for \(m\neq 0\) can be treated as a small parameter element-wise. The effect of illumination is captured by the difference of Floquet effective Hamiltonian \(H_{\text{Floq}}\) and the original time-independent Hamiltonian \(H_0\),

\begin{equation}
    \Delta H = H_{\text{Floq}} - H_0.
\end{equation}

This extra light-generated contribution \(\Delta H\) can potentially modify the symmetries originally present in the time-independent parent Hamiltonian and thus can change the band structure and other topological properties significantly. This is the essence of Floquet band engineering. In our BCL case, we have the tunability in terms of the parameters \(\eta,r\) and \(\alpha\), which will appear in the drive-induced term \(\Delta H\). As we shall see next, the interplay among these parameters gives significant control over modifying the band structures of parent multifold fermion and line-node semimetals.

\begin{figure}[t]
    \centering
    \includegraphics[width=0.48\linewidth]{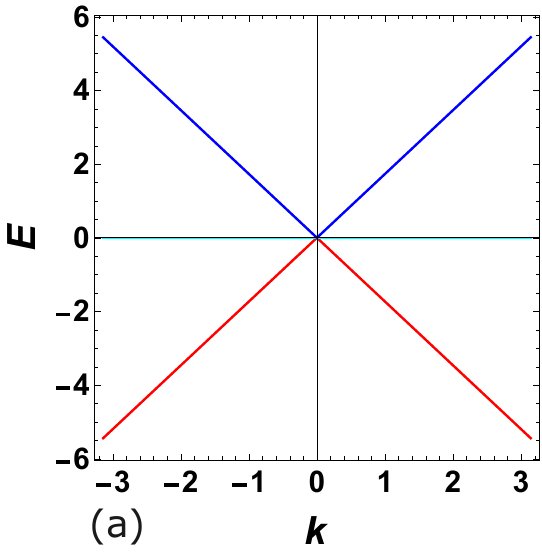}
    \includegraphics[width=0.48\linewidth]{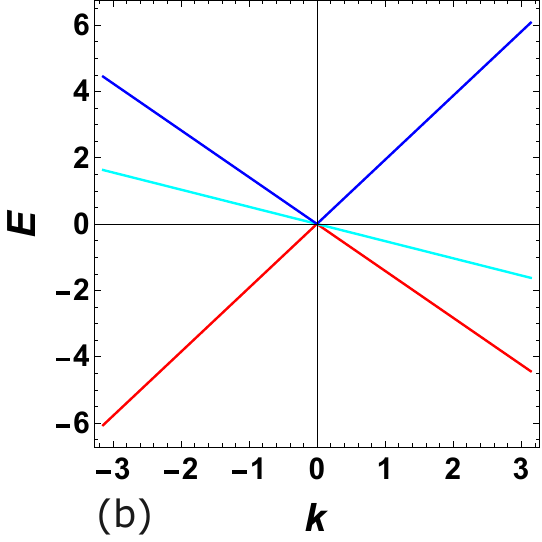}
    \caption{\textbf{Bandstructure of triplefold fermion model.} The band-structure of the effective low-energy Hamiltonian of triplefold fermion along the line \(k_x=k_y=k_z=k\), for different values of the parameter \(\phi\) with (a) \(\phi=\pi/6\)  and (b) \(\phi=\pi/12\). The band-touching point is at \(k_x=k_y=k_z=0\).}
    \label{fig:multifold band-structure}
\end{figure}

\section{Multifold fermions under BCL}\label{III}

In this section, we will first examine the effect of BCL illumination on multifold fermions by employing an effective low-energy model. We will subsequently extend our analysis to the lattice Hamiltonian. 

We have chosen the general low-energy Hamiltonian around a threefold degenerate point as proposed by Bradlyn \textit{et al.}~\cite{Bradlyn_2016},

\begin{equation}
H_{3f}(\mathbf k)= E_0 + \hbar v_f
\left(
\begin{array}{ccc}
 0 & e^{i \phi } k_x & e^{-i \phi } k_y \\
 e^{-i \phi } k_x & 0 & e^{i \phi } k_z \\
 e^{i \phi } k_y & e^{-i \phi } k_z & 0 \\
\end{array}
\right),
\end{equation}

where \(E_0\) is an energy offset, \(v_f\) is the effective velocity around the band-touching point \(\mathbf k = 0\), and $\phi$ is a material-dependent parameter. We can consider the three-band touching point to be at zero energy and thus set $E_0 = 0$. We also work in the units \(\hbar=1\) and \(v_f = 1\). Diagonalizing the Hamiltonian yields three eigenvalues $ E = 0, \pm k$ for $\phi = \pi/6$ (modulo $\pi/3$). The band structure has complete rotational symmetry about the degeneracy at $\mathbf k = 0$ for $\phi = \pi/6$ (modulo $\pi/3$), as shown in Fig.~\ref{fig:multifold band-structure}(a). Away from these $\phi$ values, the conical bands tilt (see Fig.~\ref{fig:multifold band-structure}(b)). Furthermore, the tilt is maximum for $\phi = 0$ (modulo $\pi/3$).

Let us next introduce the BCL incident on this multifold fermion system. We choose the BCL vector potential as,

\begin{equation}
    \begin{aligned}
        A_x = \mathcal A_0 \left(r \cos (\alpha -\eta  \omega  t)+\cos (\omega  t)\right),\\
        A_y = \mathcal A_0 \left(r \sin (\alpha -\eta  \omega  t)+\sin (\omega  t)\right).
    \end{aligned}
\end{equation}

Following a Peierls substitution, taking the charge to be \(e=1\), we find \( H(\mathbf k,t) = H_{3f}(\mathbf k + \mathbf A(t))\) and compute the Fourier modes. In our case, with relative frequency ratio $\eta$, not all the Fourier components will contribute. It turns out that only the terms $H_{\pm 1}$, $H_{\pm \eta}$ are finite. Using the Floquet-Magnus expansion, we can write the effective Hamiltonian as follows,

\begin{equation}
  \begin{aligned}
    H_{eff} &\approx H_0 + \frac{[H_{+1},H_{-1}]}{\omega} + \frac{[H_{+\eta},H_{-\eta}]}{\eta \omega} \\
    &+ \frac{1}{\omega} \frac{[H_0,H_{+1}]-[H_0,H_{-1}]}{2} \\
    &+ \frac{1}{\eta\omega} \frac{[H_0,H_{+\eta}]-[H_0,H_{-\eta}]}{2} +\mathcal{O}\left(\frac{1}{\omega^2}\right),\\
    &= H_{3f} + \Delta H.
  \end{aligned}
\end{equation}

\subsection{A. Results for effective low-energy model}

\begin{figure}
    \centering
    \includegraphics[width=0.9\linewidth]{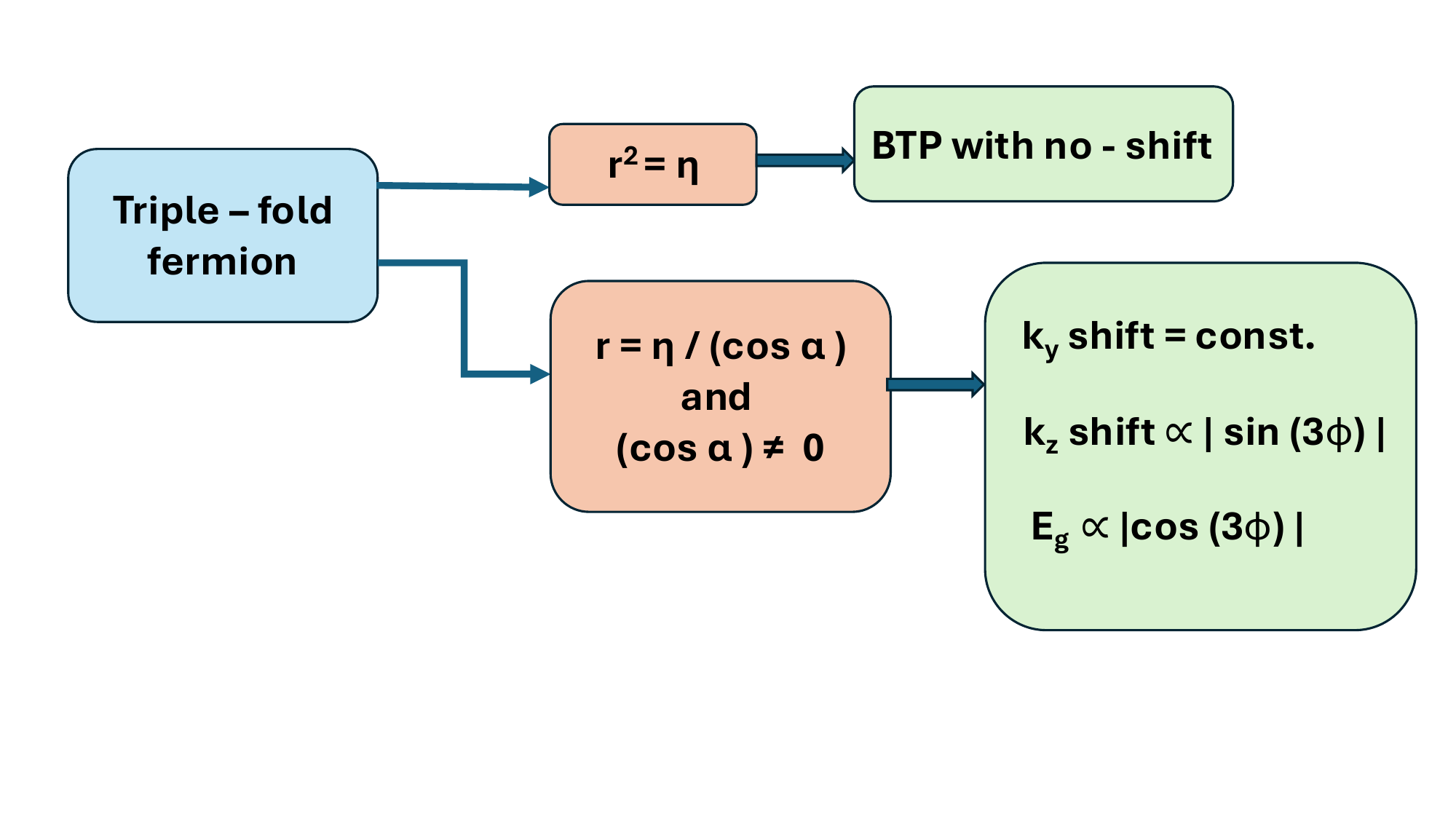}
    \caption{\textbf{Summary of effect of BCL illumination on multifold fermions.} Flow chart summarizes the tunable band structures obtained from multifold fermions depending on the different parameters of the BCL.}
    \label{fig: multifold scheme}
\end{figure}

For the effective low-energy model of the triplefold fermion we can carry out the Floquet-Magnus expansion and evaluate the effect of BCL illumination analytically. The key results are summarized in the Fig.~\ref{fig: multifold scheme}.

We find that the non-zero Fourier components of the Peierls substituted time-dependent Hamiltonian that contribute are,

\begin{widetext}
\begin{equation}
        H_{+1}= \mathcal A_0
\begin{pmatrix}
 0 & e^{i \phi }/2 & -i e^{-i\phi }/2 \\
e^{-i \phi }/2 & 0 & 0 \\
 -i e^{i\phi}/2 & 0 & 0
   \\
\end{pmatrix},
\,\,\,
 H_{-1} = \mathcal A_0 \begin{pmatrix}
 0 & e^{i \phi }/2 &  i e^{-i\phi}/2 \\
 e^{-i \phi }/2 & 0 & 0 \\
  i e^{i \phi }/2 & 0 & 0 \\
\end{pmatrix}.   
\end{equation}

\begin{equation}
        H_{+\eta} = \mathcal A_0 r \begin{pmatrix}
            0 & e^{i(\phi-\alpha)}/2 & i e^{-i(\phi+\alpha)}/2 \\
            e^{-i(\phi+\alpha)}/2 & 0 & 0\\
            i e^{i(\phi-\alpha)}/2 & 0 & 0
        \end{pmatrix},\,\,\,
        H_{-\eta} = \mathcal A_0 r \begin{pmatrix}
            0 & e^{i(\phi+\alpha)}/2 & -i e^{-i(\phi-\alpha)}/2 \\
            e^{-i(\phi-\alpha)}/2 & 0 & 0\\
            -i e^{i(\phi+\alpha)}/2 & 0 & 0
        \end{pmatrix}.
\end{equation}
\end{widetext}

Using the above expressions, we obtain \(\Delta H\) as,

\begin{equation}
    \Delta H = \begin{pmatrix}
        0 & a_1(\mathbf k) & a_2(\mathbf k) \\
        a_1(\mathbf k)^* & 0 & a_3(\mathbf k)\\
        a_2(\mathbf k)^* & a_3(\mathbf k)^* & 0
    \end{pmatrix},
\end{equation}

with,

\begin{equation}
    \begin{aligned}
        a_1(\mathbf k) &= \frac{i e^{-2i\phi}k_z}{2\omega\eta}\mathcal A_0(\eta - r \cos \alpha),\\
        a_2 (\mathbf k) &= \frac{i e^{2i\phi}k_z}{2\omega\eta} \mathcal A_0 r\sin\alpha, \\
        a_3 (\mathbf k) &= - \frac{i e^{-2i\phi}}{2\omega\eta}\mathcal A_0(\mathcal A_0 (r^2-\eta) \\
        &+ k_x(\eta - r\cos\alpha) - r k_y\sin\alpha).
    \end{aligned}
\end{equation}

\begin{figure}[t]
    \centering
    \includegraphics[width=0.45\linewidth]{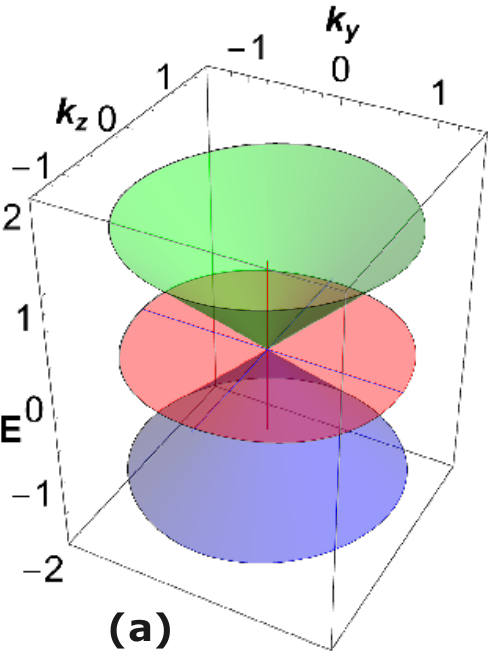}
    \includegraphics[width=0.45\linewidth]{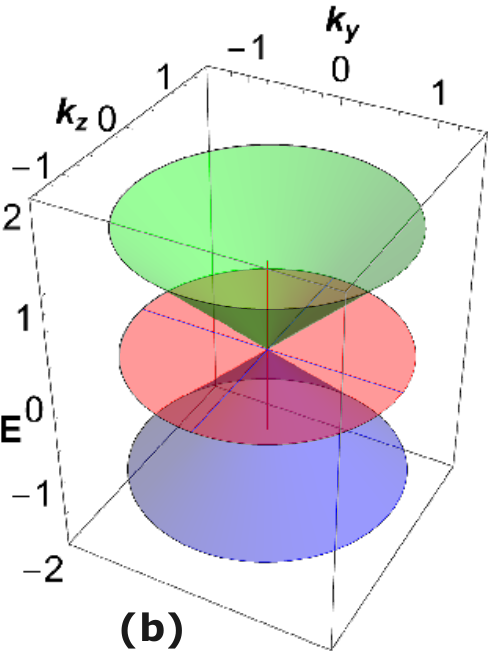}
    \vskip\baselineskip
    \includegraphics[width=0.45\linewidth]{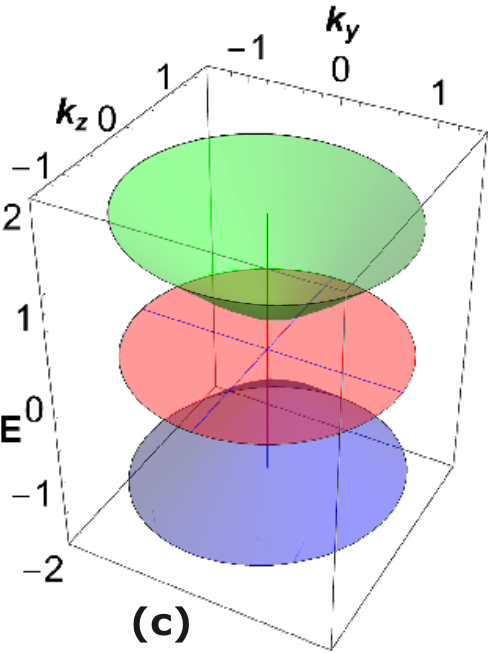}
    \includegraphics[width=0.45\linewidth]{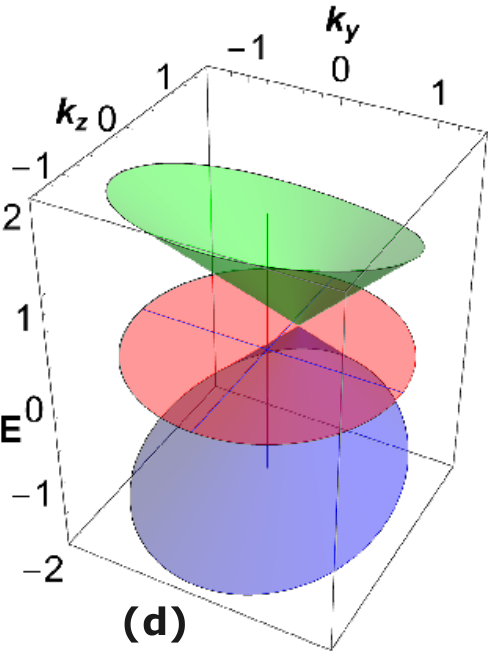}
    \caption{\textbf{Effect of BCL on low-energy model of triplefold fermions.} The amplitude ratio \(r\) and relative phase \(\alpha\) of the BCL allow tuning of the bandstructure of triplefold fermions. Top: \(r^2=\eta=2\) and \(\alpha=\pi/2\) with (a) \(\phi=0\), and (b) \(\phi=\pi/6\). We observe no shift or no band gap opening in both cases. Bottom: \(r\cos\alpha=\eta=2\) and \(\alpha=\pi/4\) with (c) \(\phi=0\), and (d) \(\phi=\pi/6\). We observe band gap opening for \(\phi=0\) and the band touching point shift for \(\phi=\pi/6\).}
    \label{fig: bcl on multifold analytical}
\end{figure}

\begin{figure}[t]
    \centering
    \includegraphics[width=0.97\linewidth]{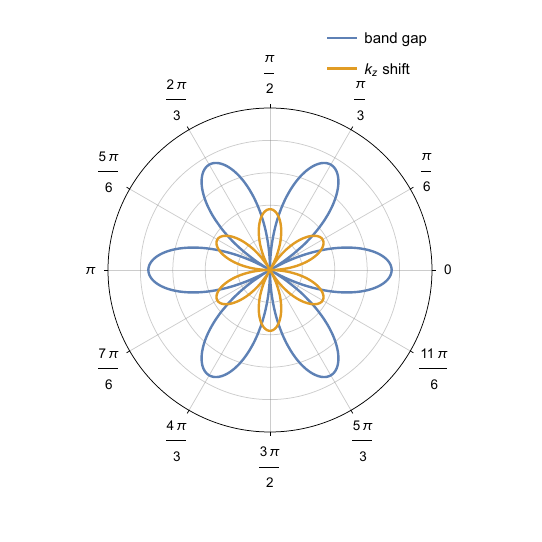}
    \caption{\textbf{Variation of band gap and band touching point shift with \(\phi\)}. For the case \(r=\eta/\cos\alpha\), the band gap and the magnitude of the shift of the band crossing point along \(k_z\) take a periodic form as a function of the parameter \(\phi\). The periodicity is \(\pi/3\) as can be seen from the plot.}
    \label{fig: polar plot}
\end{figure}

Now to understand the effects of the BCL, we focus on the analytical form of $\Delta H$. From the above three expressions of $a_1$, $a_2$ and $a_3$, we can analyze their different behaviours. Let next us consider the different features arising from the various choices of the parameters.

For the case \(r=\eta,\,\alpha=0\), we find that both $a_1$, $a_2$ are zero and $a_3$ is a constant. This is exactly the same, up to a constant factor, with the light-induced Hamiltonian in Ref.~\cite{RimikaTripleFold}. The Floquet Hamiltonian has all the symmetries of $H_0$ for \(\phi=\pi/6(\text{mod } \pi/3)\). Here, for $\phi = 0$ (modulo $\pi / 3 $), the bad gap opens up, and for $\phi = \pi / 6$ (modulo $\pi / 3 $) the bands remain gapless. However, there is a shift of the band crossing point along the $k_z$ direction. Therefore, for \(r=\eta,\,\alpha=0\), the effect of BCL is similar to that of the single-frequency light, i.e., \(r=0\).

Remarkably, we find that the choice \(r^2 = \eta\) always preserves the band-touching point and the touching point remains at \(k_x=k_y=k_z=0\), irrespective of the values of \(\phi\) and \(\alpha\). This can be understood as \(H_{eff}\) vanishes at \(k_x=k_y=k_z=0\), such that all the eigenvalues also vanish at this point. As an example, we choose $\eta = 2$, $\alpha = \pi / 2$, and plot the bands for both $\phi = 0$ and $\phi = \pi / 6$. The resulting band structures are shown in Fig.~\ref{fig: bcl on multifold analytical}(a) and Fig.~\ref{fig: bcl on multifold analytical}(b).

We find that for \(\eta = r \cos\alpha\), such that $\cos{\alpha} \neq 0 $ and \(k_x=0\), the constant term in the eigenvalue equation of $H_{eff}$ is zero. This indicates that one eigenvalue is zero for all \(k_y\) and \(k_z\). Therefore, this choice of BCL parameters allows engineering a complete flat band in the $k_y-k_z$ plane. The other two bands are gapped in this plane. We find that the band gap becomes minimal at

\begin{equation}
\begin{aligned}
     &k_y = \frac{\mathcal A_0^3 \tan\alpha (\eta \sec^2\alpha -1)}{4\omega^2 + \mathcal A_0^2 \tan^2\alpha},\\
     &k_z = \frac{2\mathcal A_0^2 \omega \sin 3\phi (\eta \sec^2\alpha -1)}{4\omega^2 + \mathcal A_0^2 \tan^2\alpha},
\end{aligned}
\end{equation}

and the square of the minimum value of the band gap is,

\begin{equation}
    E_{gap}^2 = \frac{\mathcal A_0^4 (1+\cos 2\alpha - 2\eta)^2 \cos^2 3\phi\,\sec^4 \alpha }{4\omega^2 + \mathcal A_0^2 \tan^2\alpha}.
\end{equation}

We note that the band gap vanishes for \(\phi = \pi/6 (\text{mod } \pi/3)\), as shown in  Fig.~\ref{fig: polar plot}. This results in a three-fold band touching point. The corresponding band structure plots for \(\alpha=\pi/4\) are presented in Fig.~\ref{fig: bcl on multifold analytical}(c) and Fig.~\ref{fig: bcl on multifold analytical}(d).

Finally, the choice \(\alpha=\pi/2\) always opens up a gap in the \(k_y-k_z\) plane unless \(r^2=\eta\). We observe a large gap with a small shift for \(\phi=0\) and the small gap with a large shift for \(\phi=\pi/6\), but the gap does not vanish.

Overall, we find that BCL allows remarkable control over the band structure of triplefold fermion systems. By tuning the BCL parameters \(r\), $\eta$, and \(\alpha\) we can tune the band-gap and the shift, in a model-independent way. For the single frequency case, the values of both the band-gap and the shift depends on the model dependent parameter \(\phi\)~\cite{RimikaTripleFold}. Therefore, at generic values of \(\phi\), both the band-gap and the shift will be non-zero. In contrast, with BCL, we have found that for \(r^2=\eta\), band-gap and shift are both vanishing for any value of \(\phi\). Hence, even for generic values of \(\phi\), we can tune the system from gapped to gapless or vice-versa by changing only \(r\), which is not possible for the single frequency case. This directly highlights the versatility of BCL in tuning the properties of topological semimetals.

\subsection{B. Lattice model for triplefold fermions}

\begin{figure*}[t]
    \centering
    \includegraphics[width=0.35\linewidth]{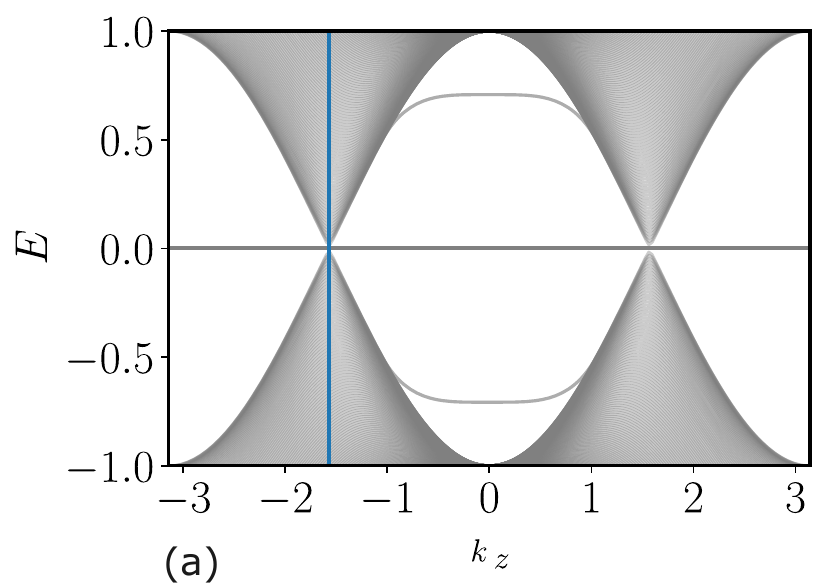}
    \includegraphics[width=0.35\linewidth]{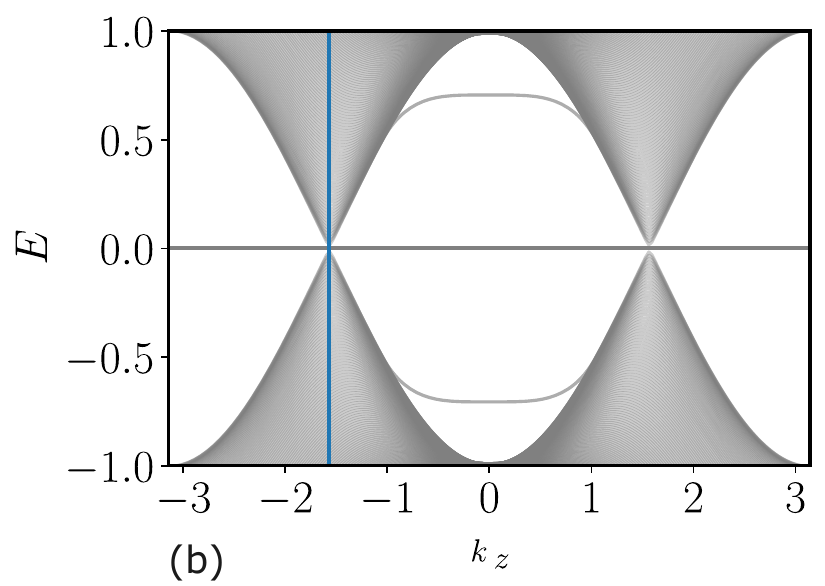}
    \vskip\baselineskip
    \includegraphics[width=0.35\linewidth]{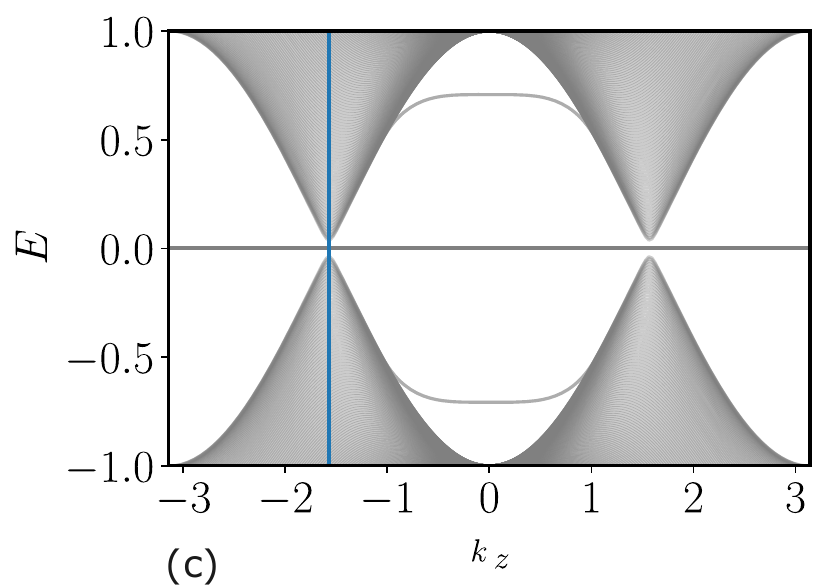}
    \includegraphics[width=0.35\linewidth]{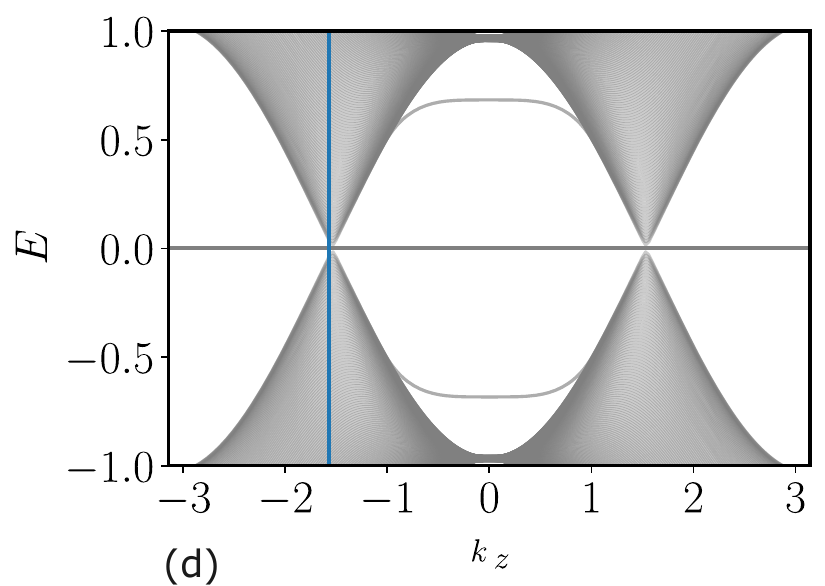}
    \caption{\textbf{Effect of BCL on triplefold fermion lattice model.} The band structures as a function of \(k_z\) with fixed \(k_y=0\) for the lattice model for different $r$ and $\phi$. Here we have used open boundary conditions along the $x$ direction (with 200 sites), while along $y$ and $z$ we have periodic boundary conditions. Top-panel: Band structure for relative amplitude $ r = \sqrt{2}$ for the case \(r^2=\eta\) with (a) \(\phi=0\) and (b) \(\phi=\pi/6\). We do not observe any band gap or shift of the band touching point for these cases. Bottom panel: For \(r=\eta\), we find a band gap opening for (c) \(\phi=0\), and shift of the band touching point for (d) \(\phi=\pi/6\). The blue vertical line corresponds to \(k_z=\pi/2\). Here we have set \(\eta=2\), \(\alpha=0\), \(\mathcal A_0 = 0.5\), \(\omega=3\), and \(m=t=1\). These numerical results confirm the analytical predictions from the effective low-energy model.}
    \label{fig: Multifold numerical}
\end{figure*}

To further explore the effects of BCL under open boundary conditions, we consider a lattice model that has two triply degenerate band touching points. The Hamiltonian reads~\cite{RimikaTripleFold}

\begin{widetext}
\begin{equation}\label{eq: triplefold lattice H}
    H_{tb}(\mathbf k) =
    \begin{pmatrix}
        0 & e^{i\phi}t\sin k_x & e^{-i\phi}t\sin k_y \\
        e^{-i\phi}t\sin k_x & 0 & e^{i\phi} m(\cos k_z+\cos k_x + \cos k_y -2)\\
        e^{i\phi}t\sin k_y & e^{-i\phi} m(\cos k_z+\cos k_x + \cos k_y -2) & 0
    \end{pmatrix},
\end{equation}
\end{widetext}

where \(t\) is the hopping and \(m\) is the mass term. We consider open boundary conditions along the \(x\)-direction and numerically compute the band structures.

In Fig.~\ref{fig: Multifold numerical} we show the numerically calculated band structures under open boundary conditions when BCL is applied. Fig.~\ref{fig: Multifold numerical}(a) and Fig.~\ref{fig: Multifold numerical}(b) show that for \(r^2=\eta\) the band-touching point remains intact, with no shift for any chosen value of \(\phi\). On the other hand, if we have \(r=\eta\) (for \(\alpha=0\)), then we find that there is a band gap opening with no shift for \(\phi=0\) (see Fig.~\ref{fig: Multifold numerical}(c)), and there is no band gap but a shift in BTP for \(\phi=\pi/6\) (see Fig.~\ref{fig: Multifold numerical}(d)). These findings our fully consistent with our results from the low-energy model.

Furthermore, our multifold fermion model has non-trivial topology and therefore it has topologically protected surface states in the slab geometry, arising from the bulk-boundary correspondence. From Fig.~\ref{fig: Multifold numerical} we observe that the surface states remain intact even in the presence of BCL, and they connect the shifted band touching points when there is a shift. This further elucidates the non-trivial topology under illumination with BCL.

\section{Line-node semimetals under BCL}

\begin{figure}
    \centering
    \includegraphics[width=0.9\linewidth]{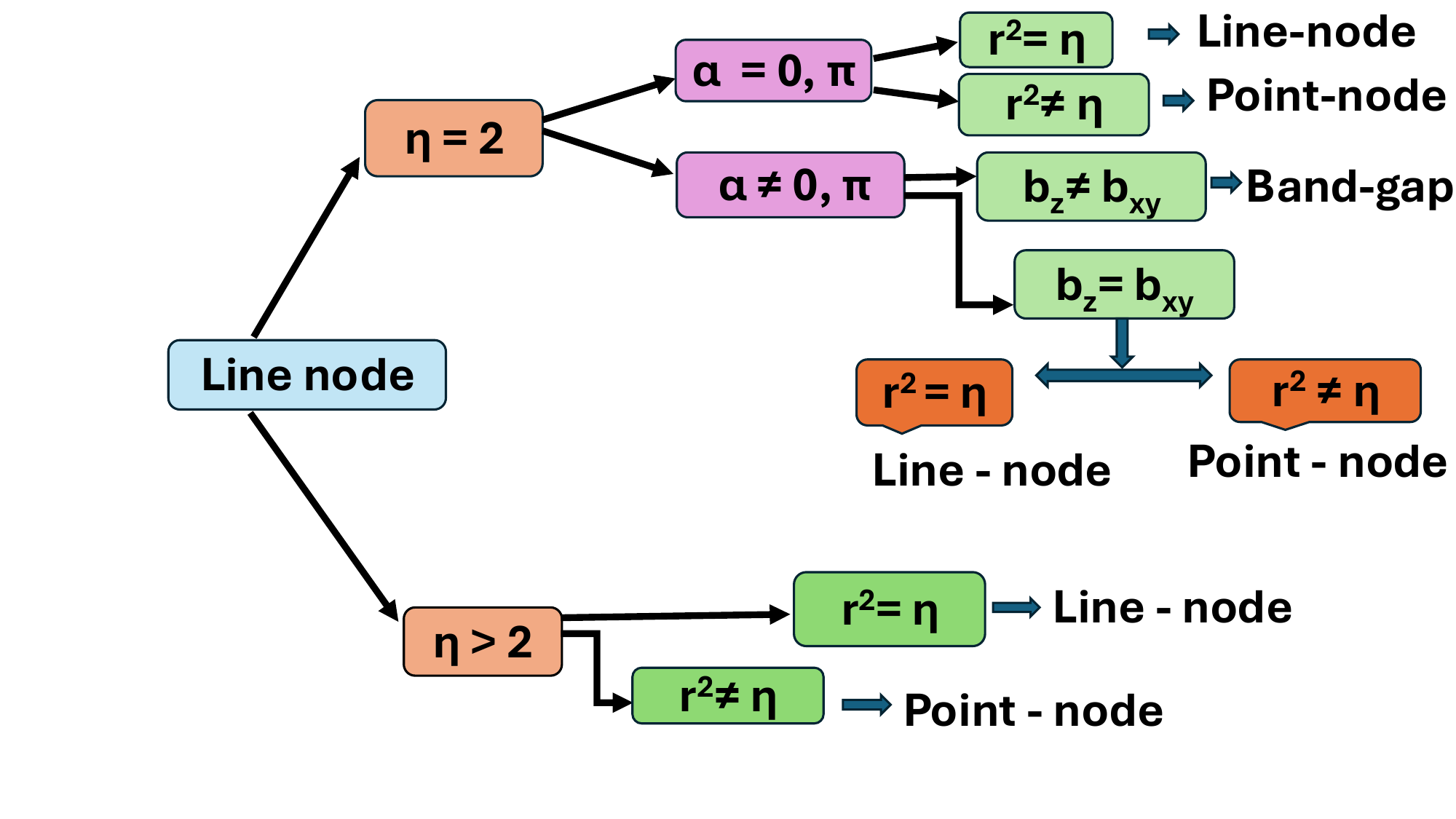}
    \caption{\textbf{Summary of effect of illumination of BCL on line-node semimetals.} Flow chart summarizes the tunable band structures obtained from line node semimetals depending on the different parameters of the BCL.}
    \label{fig: line-node scheme}
\end{figure}

In this section, we next present our findings on the effects of BCL on line-node semimetals. These are a class of topological semimetals whose bands touch along a line instead of points \cite{ln1,ln2,ln3,ln4,Gao_2019}. In previous studies, it was shown that the application of circularly or elliptically polarized light could drive a Lifshitz-like transition in line-node semimetals, giving rise to point nodes~\cite{ANLineNodeCL,Chan_2016,PhysRevLett.117.087402}. Here, we take the idea forward and explore the effects of the application of BCL on these topological semimetals. As we show next, BCL offers significantly more control with greater tunability and richer phases arising from line-node semimetals. In Fig.~\ref{fig: line-node scheme}, we schematically show the effect of BCL on line-node semi-metals, which will be obtained analytically in this section.

To start with, we analyze the the low-energy Hamiltonian for line-node semimetals, which reads~\cite{LineNodeKimetal},

\begin{equation}\label{eq: line-node eff H}
\begin{aligned}
    H &= [\epsilon_0 + a_{x y} (k_x^2 + k_y^2) + a_z k_z^2] I + v k_z \sigma_y \\
    &+ [\Delta \epsilon + b_{xy}(k_x^2 + k_y^2) + b_z k_z^2] \sigma_z,
\end{aligned}
\end{equation}

where \(\sigma_x,\sigma_y,\sigma_z\) are the Pauli matrices and \(I\) is the 2\(\times\)2 identity matrix. Here \(\epsilon_0,a_{xy},a_z,v,\Delta\epsilon,b_{xy},b_z\) are the system parameters in the effective model.

The energy eigenvalues for this Hamiltonian are found to be,

\begin{equation}\label{lndis}
\begin{aligned}
    E &= [\epsilon_0 + a_{x y} (k_x^2 + k_y^2) + a_z k_z^2] \\
    &\pm \sqrt{[\Delta \epsilon + b_{xy}(k_x^2 + k_y^2) + b_z k_z^2]^2 + v^2 k_z^2}.
\end{aligned}
\end{equation}

\begin{figure}[t]
    \centering
    \includegraphics[width=0.5\linewidth]{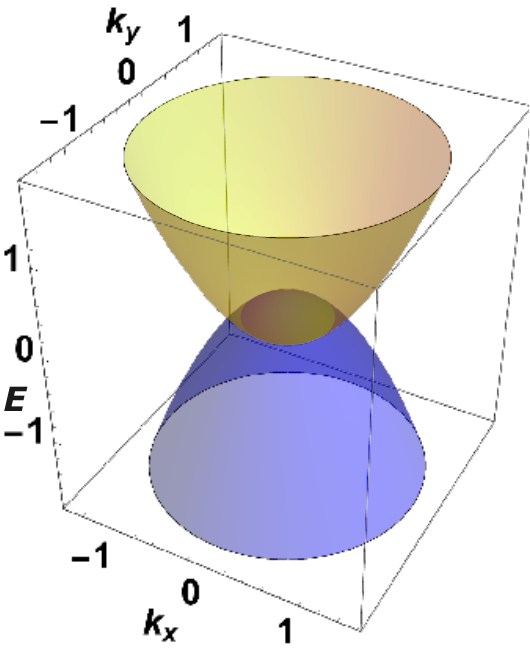}
    \caption{\textbf{Line-node semi-metal band structure.} The band structure for the line-node semi-metal (in the \(k_z=0\) plane) before application of light to the system. The chosen parameters are \(\epsilon_0=0,a_{xy}=0,a_z=1,b_{xy}=1,b_z=1,v=1\) and \(\Delta\epsilon=-0.2\).}
    \label{fig: line-node diag}
\end{figure}

Thus, it is possible to obtain a line nodal dispersion for \(\Delta \epsilon <0\), as shown in Fig.~\ref{fig: line-node diag}. From Eq.~\ref{lndis} it can be seen that the bandgap vanishes along the circle \(k_x^2 + k_y^2 = - \Delta \epsilon/b_{xy}\) for \(k_z = 0\). To explore the effects of BCL further, we consider the vector potential of the form \(\vec A = A_y \hat y + A_z \hat z\) with,

\begin{equation}
\begin{aligned}
    &A_y = \mathcal A_0 (- r \sin (\eta \omega t - \alpha) + \sin \omega t ),\\
    &A_z = \mathcal A_0 (+ r \cos (\eta \omega t - \alpha) + \cos \omega t).
\end{aligned}
\end{equation}

Here all the parameters in the vector potential are the same as already mentioned in Sec.~\ref{III}. 

\subsection{A. Results for low-energy model}

\begin{figure}[t]
    \centering
    \includegraphics[width=0.45\linewidth]{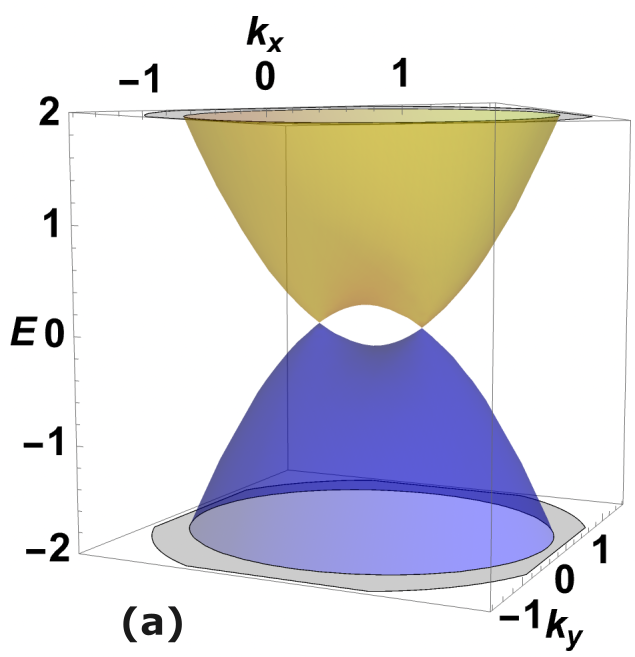}
    \includegraphics[width=0.45\linewidth]{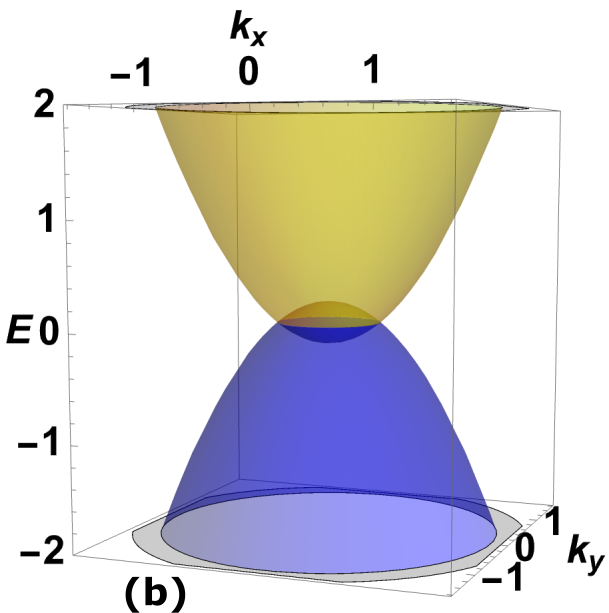}
    \vskip\baselineskip
    \includegraphics[width=0.45\linewidth]{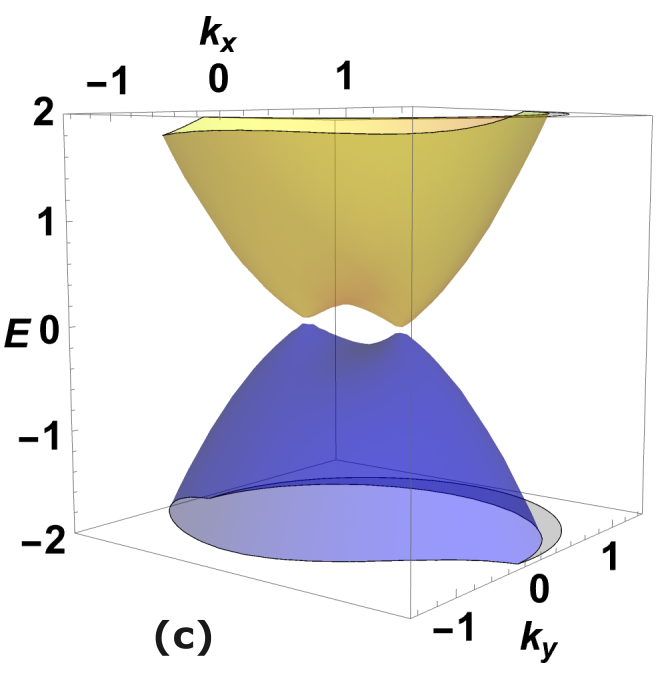}
    \includegraphics[width=0.45\linewidth]{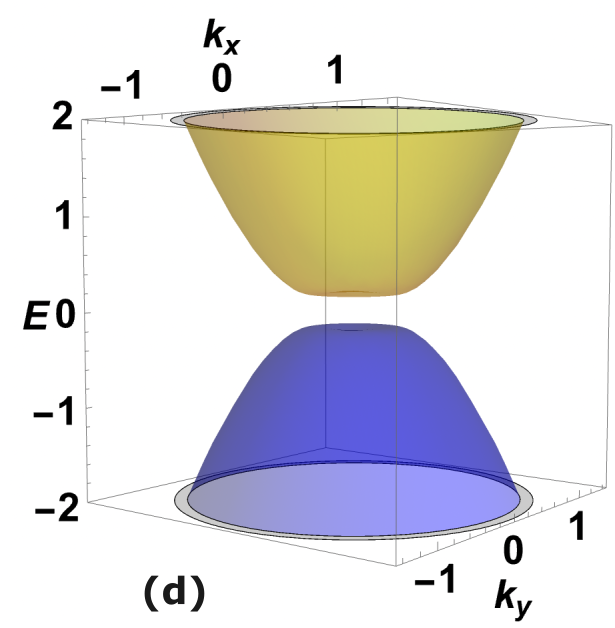}
    \vskip\baselineskip
   \includegraphics[width=0.45\linewidth]{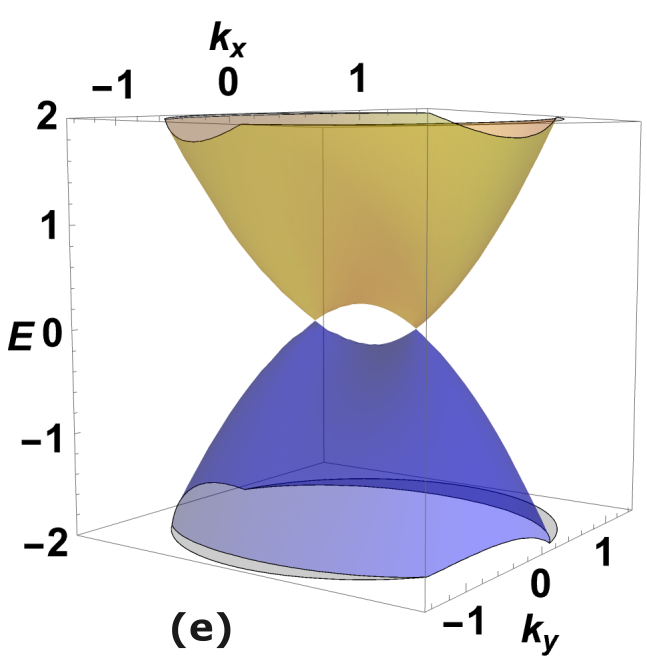}
   \includegraphics[width=0.45\linewidth]{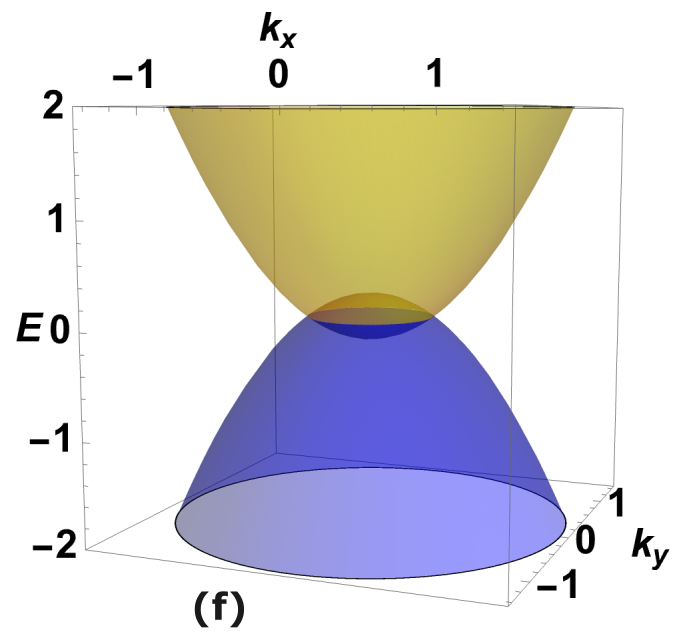}
   \caption{\textbf{Effect of BCL on line-node semimetal.} The band structure of line node semimetal under illumination with BCL choosing \(b_{xy}\neq b_z\). Top-panel: Effect of BCL with \(\eta=2\), \(\alpha=0\) while keeping (a) \(r=1\) and (b) \(r=\sqrt{\eta}=\sqrt{2}\). Middle-panel: Effect of BCL with \(\eta=2\), \(\alpha=\pi/2\) choosing (c) \(r=1\) and (d) \(r=\sqrt{\eta}=\sqrt{2}\). Bottom-panel: Effect of BCL with \(\eta=4\), \(\alpha=\pi/2\) and selecting (e) \(r=1\) and (f) \(r=\sqrt{\eta}=2\). Therefore, the phase difference \(\alpha\) between the two CLs present in BCL plays an important role in determining the light-tuned band structure and the precise effect depends on the frequency ratio \(\eta\).
   The values of other parameters are \(\epsilon_0=0,a_{xy}=0,a_z=1,v=1,b_{xy}=1,b_z=1.2,\Delta\epsilon=-0.2,\omega=1,\mathcal A_0 =1\).}
   \label{fig: effect of BCL line-node analytical}
\end{figure}

We analyze the effect of illumination of the above BCL vector potential on the line-node band structure using Floquet theory. We first consider the case of \(\eta = 2\). The effective Hamiltonian in the large frequency limit is obtained to be

\begin{equation}
\begin{aligned}
    H_{eff} &\approx H + \sum_{n=1}^4\left( \frac{[H_{+n},H_{-n}]}{n\omega}+ \frac{1}{n\omega} \frac{[H,H_{+n}]-[H,H_{-n}]}{2}\right), \\
    &= H + \Delta H.
\end{aligned}
\end{equation}

The light-induced term, \(\Delta H\), turns out to be of the form \(c_x \sigma_x\), such that,

\begin{equation}
\begin{aligned}
    c_x &= \frac{v}{4\omega} [4 \mathcal A_0^2 b_{xy} k_y (r^2 - 2) - \mathcal A_0 r \sin \alpha \{3\mathcal A_0^2(b_z-b_{xy}) \\
    &+ 2 b_{xy}(k_x^2 + k_y^2 )+2 \Delta\epsilon\}]+ k_z c',
\end{aligned}
\end{equation}

where the term \(c'\) does not influence the band structure in the \(k_z=0\) plane since it multiplies \(k_z\). It takes the form,

\begin{equation}
\begin{aligned}
    c' &= \frac{\mathcal A_0 v}{12\omega} (-12 b_{xy}k_y (-2+r\cos\alpha) + r\sin\alpha (-8\mathcal A_0 (b_{xy}-b_z) \\
    &+ 6b_zk_z + 3 \mathcal A_0 r\cos\alpha (b_z-b_{xy}))).
\end{aligned}
\end{equation}

The expression for the band gap is then found to be,

\begin{equation}
    E_{gap} = 2 \sqrt{[\Delta \epsilon + b_{xy}(k_x^2 + k_y^2) + b_z k_z^2]^2 + v^2 k_z^2 + c_x^2}.
\end{equation}

For the \(k_z = 0\) plane, the expression for \(c_x\) reduces to,

\begin{equation}
\begin{aligned}
    c_x &= \frac{v}{4\omega} [4 \mathcal A_0^2 b_{xy} k_y (r^2 - 2) -\mathcal A_0 r \sin \alpha \{3\mathcal A_0^2(b_z-b_{xy}) \\
    &+ 2 b_{xy}(k_x^2 + k_y^2 )+2 \Delta\epsilon\}].
\end{aligned}
\end{equation}

Moreover, for the special cases of \(\alpha = 0 \text{ or } \pi\), \(c_x\) simplifies to,

\begin{equation}
    c_x = \frac{\mathcal A_0^2 v}{\omega} [b_{xy} (r^2 - 2)] k_y.
\end{equation}
    
Interestingly, this is analogous to the effect of single-frequency light with an extra tunability due to the amplitude ratio, \(r\). For \(r\neq \sqrt{2}\), we obtain two band touching points, as shown in Fig.~\ref{fig: effect of BCL line-node analytical}(a). On the other hand, for \(r=\sqrt{2}\), the effect of illumination is effectively nullified. We recover the line-node semimetal dispersion even in the presence of illumination, as we present in Fig.~\ref{fig: effect of BCL line-node analytical}(b). An important point to note here is that this conclusion is independent on whether \(b_{xy}\) is equal to \(b_z\) or not. 

If we have \(\alpha \neq 0 \text{ or } \pi\), then the second term in \(c_x\) does not vanish. This will lead to two cases, as follows. First, if \(b_{xy} = b_z\), then the situation turns out to be similar to the \(\alpha = 0\) case because of the combination \(2 b_{xy}(k_x^2 + k_y^2 )+2 \Delta\epsilon\). The same factor appears in the coefficient of \(\sigma_z\) in \(H\). For \(r\neq \sqrt{2}\), we obtain two band touching points, and for \(r=\sqrt{2}\), we recover the line-node dispersion. Secondly, if \(b_{xy} \neq b_z\), then for any non-zero \(r\) a band gap opens up. This results in the absence of any band touching point, as we show in Fig.~\ref{fig: effect of BCL line-node analytical}(c) and Fig.~\ref{fig: effect of BCL line-node analytical}(d).

From our above results, we can propose a direct approach to extract the anisotropy of the line nodal semimetal dispersion. If one can observe the above features of the model with the illumination of the BCL with \(\eta=2\) one will be able to conclude whether \(b_z = b_{xy}\) or \(b_z \neq b_{xy}\). This will give the information about the anisotropy of the model along the $k_z$ direction.

Let us next consider the case with \(\eta > 2\). In the \(k_z = 0\) plane, we have \(\Delta H = c_x \sigma_x\). We find that for \(\eta = 3\), 

\begin{equation}
    c_x = \frac{v[2\mathcal A_0^2 b_{xy}k_y (-3+r^2)-\mathcal A_0 r\sin\alpha\{b_{xy}(k_x^2+k_y^2)+\Delta\epsilon\}]}{3 \omega},
\end{equation}

while for \(\eta = 4\),

\begin{equation}
    c_x = \frac{v[2 \mathcal A_0^2 b_{xy}k_y (-4+r^2)-\mathcal A_0 r\sin\alpha\{b_{xy}(k_x^2+k_y^2)+\Delta\epsilon\}]}{4 \omega}.
\end{equation}

Remarkably, we note that in these cases \(r=\sqrt{\eta}\) always recovers the line-node semimetal dispersion, independent of the parameters of the Hamiltonian, i.e., \(\alpha, b_{xy}, b_z\). It appears that this conclusion is true for any \(\eta > 2\). In this case, in contrast with the results obtained with $\eta = 2$, even if \(b_{xy}\neq b_z\) and \(\alpha=\pi/2\), the conclusions hold true (see Fig.~\ref{fig: effect of BCL line-node analytical}(e) and Fig.~\ref{fig: effect of BCL line-node analytical}(f)). Thus $\eta > 2$ is strikingly different from the \(\eta=2\) case, where a gap opens up for \(b_{xy}\neq b_z\) and \(\alpha=\pi/2\).

\subsection{B. Analysis with lattice model}

\begin{figure}[t]
    \centering
    \includegraphics[width=0.47\linewidth]{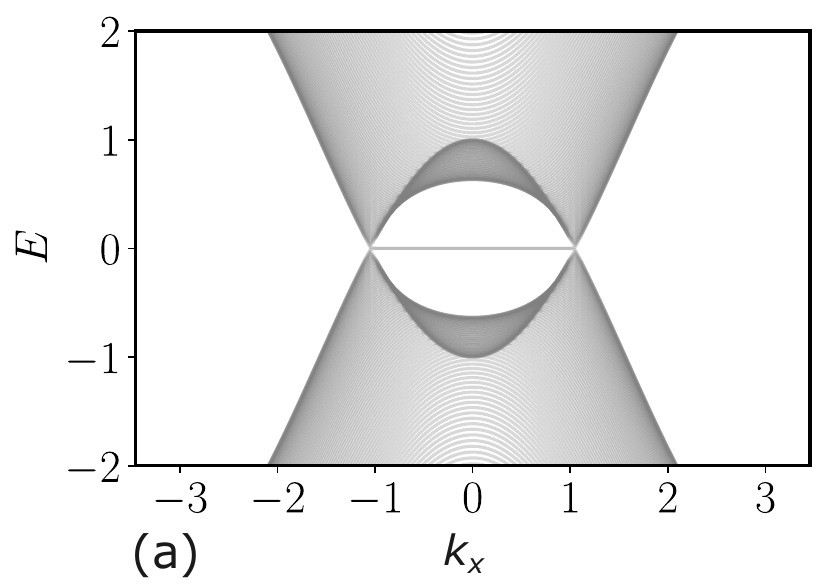}
    \includegraphics[width=0.47\linewidth]{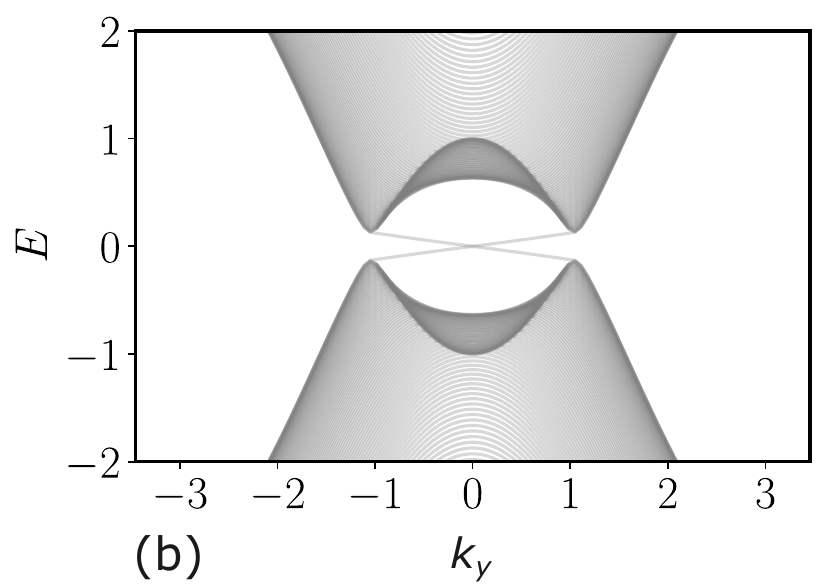}
    \vskip\baselineskip
    \includegraphics[width=0.47\linewidth]{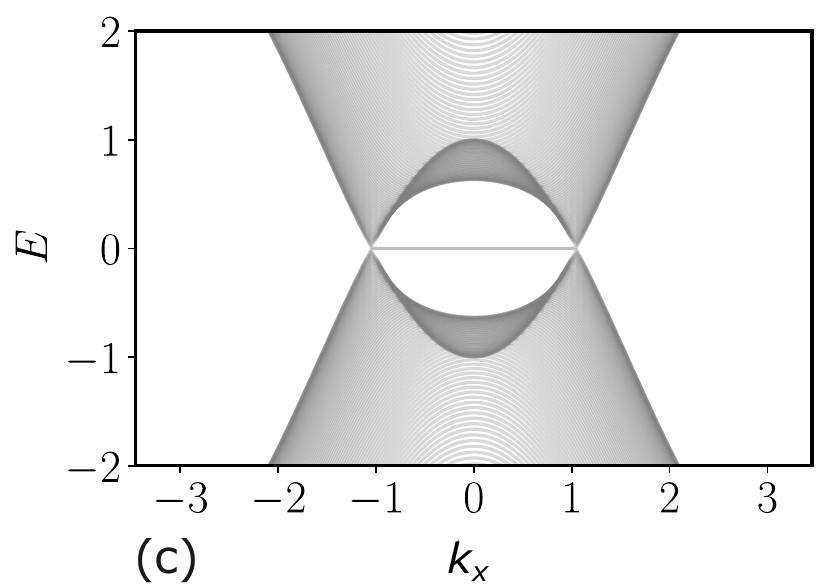}
    \includegraphics[width=0.47\linewidth]{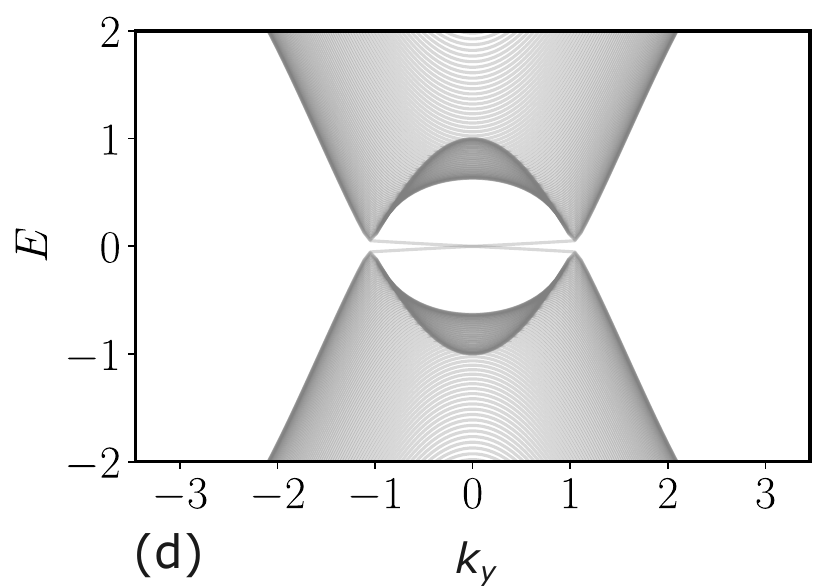}
    \caption{\textbf{Effect of BCL with \(\eta=2\), \(\alpha=0\) on lattice model of line-node semi-metal.} The effect of BCL (with \(\mathcal A_0 = 1.0\) and \(\omega=3\)) on a lattice model of line-node semi-metal with open boundary conditions along \(z\)-direction (with 200 sites) and periodic boundary conditions along \(x\) and \(y\)-directions. Parameter choices are \(v=1,\Delta\epsilon = -1.0,b_{xy}=1.0\) and \(b_z=2.0\). Top-panel: Band structure for \(r=1\) case along (a) \(k_x\) direction with fixed \(k_y=0\) and (b) \(k_y\) direction with fixed \(k_x=0\). We can observe that the band structure has two point nodes. Bottom-panel: Band structure for \(r=\sqrt{2}\) case along (c) \(k_x\) direction with fixed \(k_y=0\) and (d) \(k_y\) direction with fixed \(k_x=0\). We find that the line-node is restored for \(r=\sqrt{2}=\sqrt{\eta}\).}
    \label{fig: lattice line-node alpha=0}
\end{figure}

\begin{figure}[t]
    \centering
    \includegraphics[width=0.47\linewidth]{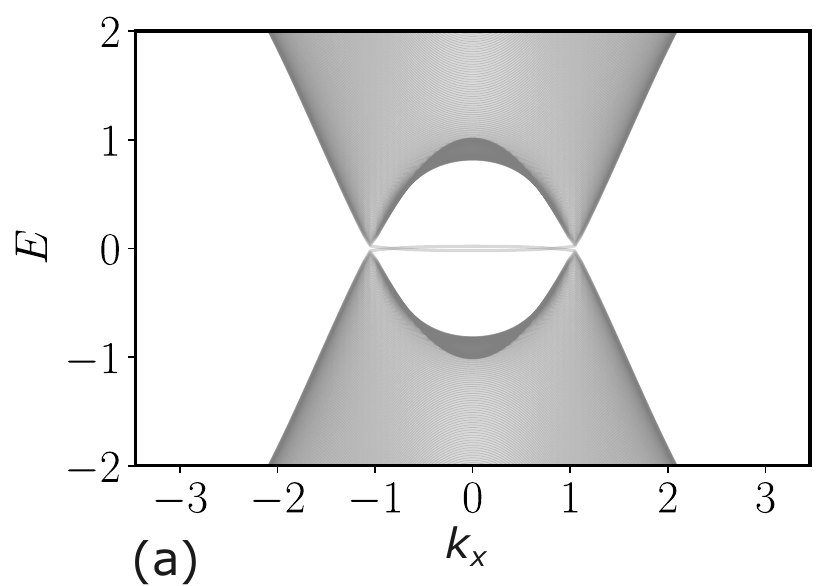}
    \includegraphics[width=0.47\linewidth]{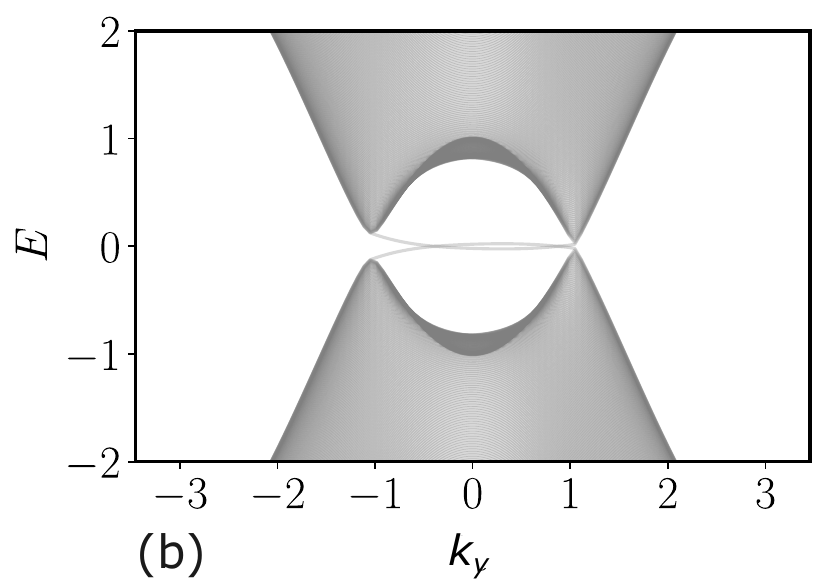}
    \vskip\baselineskip
    \includegraphics[width=0.47\linewidth]{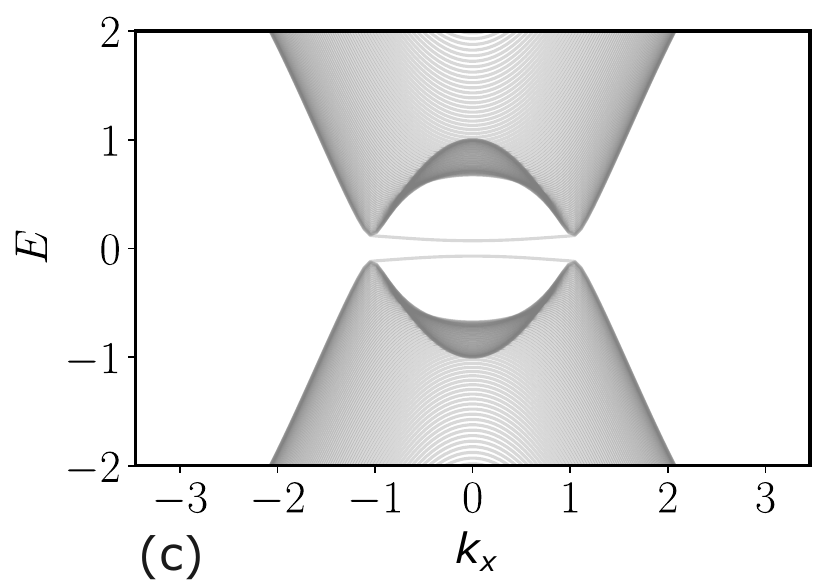}
    \includegraphics[width=0.47\linewidth]{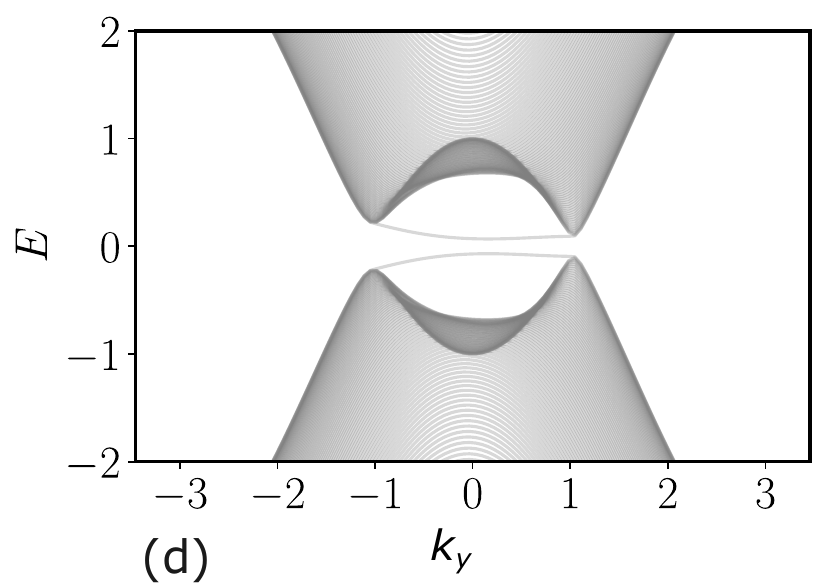}
    \caption{\textbf{Effect of BCL with \(\eta=2\), \(\alpha=\pi/2\) on lattice model of line-node semi-metal.} Here we choose \(\alpha=\pi/2,r^2=\eta=2\). Top-panel: Band structure for the case \(b_{xy}=b_z\) along (a) \(k_x\) direction with fixed \(k_y=0\) and (b) \(k_y\) direction with fixed \(k_x=0\). Bottom-panel: Band structure for the case \(b_{xy}\neq b_z\) along (c) \(k_x\) direction with fixed \(k_y=0\) and (d) \(k_y\) direction with fixed \(k_x=0\). We find that the surface states for \(\alpha=\pi/2\) have different connectivity than the cases with \(\alpha=0\). For \(b_{xy}=b_z\) case we have taken \(b_{xy}=b_z=1.0\) and for \(b_{xy}\neq b_z\) case we have taken \(b_{xy}=1.0, b_z = 2.0\). Other parameters are same as in Fig.~\ref{fig: lattice line-node alpha=0}.}
    \label{fig: lattice line-node alpha=pi/2}
\end{figure}

In the previous subsection we have analytically determined the effect of BCL illumination on an effective model of line-node semimetal. We find that BCL is a versatile tool for controlled tuning of line-node band structure as well as to infer information about the underlying model. Next, we further explore the effects of BCL on a lattice model of line-node semimetal, for which we perform a numerical analysis.

A lattice model which shows the line-node semimetallic dispersion and reduces to the previous effective Hamiltonian near \(\mathbf k \approx 0\) reads,

\begin{equation}
\begin{aligned}
    H(\mathbf k) &= v \sin k_z \sigma_y + [\Delta \epsilon - 2 b_{xy}(\cos k_x + \cos k_y -2)\\
    &- 2 b_z (\cos k_z -1)]\sigma_z.
\end{aligned}
\end{equation}

To study the effect of BCL on finite-size slabs, we use open boundary conditions in the \(z\) direction, by keeping periodic boundary conditions along $x$ and $y$ directions.

To understand the band structure as a function of \(k_x\) and \(k_y\), we have plotted the bands as a function of \(k_x\) for fixed \(k_y=0\) as well as a function of \(k_y\) for fixed \(k_x=0\). The top panel of Fig.~\ref{fig: lattice line-node alpha=0} shows the band structure for the BCL parameter choice \(\eta=2,\alpha=0,r=1\). The plots confirm that these parameter choices indeed result in two points nodes, as found from our analytical analysis on the effective low-energy model. On the other hand, the case \(r=\sqrt{2}\) restores the line-node semimetal, as shown in the bottom panel of Fig.~\ref{fig: lattice line-node alpha=0}. For these plots we choose \(b_{xy}\neq b_z\) and our findings establish that the relative values of \(b_{xy}\) and \(b_z\) do not affect the band structure qualitatively when the relative phase \(\alpha\) in BCL with \(\eta=2\) is zero.

From our analytical results of the low energy model, we found that if we choose \(\alpha=\pi/2\) for \(\eta=2\), then the relative values of \(b_{xy}\) and \(b_z\) can affect the band structure significantly. From numerical analysis on the lattice model we come to the same conclusion. The top panel of Fig.~\ref{fig: lattice line-node alpha=pi/2} shows the band structure for \(\eta=2,\alpha=\pi/2,r=\sqrt{2}\) and \(b_{xy}=b_z\), which is indeed a line node, as expected. Now, if we choose \(b_{xy}\neq b_z\) keeping the other parameters as is, then from the bottom panel of Fig.~\ref{fig: lattice line-node alpha=pi/2} we observe that the band gap opens up. Furthermore, the dispersion of the surface states indicates that the system has become a trivial insulator. This is unlike the case for \(\alpha=0\) in Fig.~\ref{fig: lattice line-node alpha=0}, where we observe the surface states to be topologically non-trivial. From the surface state plots for $\alpha = 0$ we can see that there is only one left-moving and one right-moving state, which indicate the presence of non-trivial topologically protected surface states having Chern number $1$. 

\section{Summary and Outlook}

To summarize, we have investigated the effect of illumination of BCL on multifold fermion and line node semimetals. Our analytical and numerical results unravel the remarkable tunability of topological phases by employing BCL, in contrast to single circularly polarized light. We have shown how the band structures of both multifold fermion and line-node semimetals can show rich features with the BCL illumination, by means of both low-energy and lattice models. For multifold fermions, we have shown that if the relative amplitude $r$ between the two frequency components of the BCL is related to the relative frequency $\eta$ in such a way that $r^2 = \eta$, then the band touching points remain unchanged without any shift for any choice of the material-dependent parameter $\phi$. For other choices of the BCL parameters, if we have \(r=\eta\) (for \(\alpha=0\)) then there is a band gap opening with no shift for \(\phi=0\) and there is no band gap but a shift in band-touching point for \(\phi=\pi/6\). In the case of line-node semimetal for $\eta = 2$ and $\alpha = 0, \pi $, we have shown both analytically and numerically that for $r^2 = \eta$, we always recover the line node even after the illumination of BCL. On the other hand, for $r^2 \neq \eta$, we obtain two point nodes. Strikingly, for $\eta > 2$ with $r^2 = \eta$, we always obtain the line-node dispersion independent of the model parameters. In conclusion, we are hopeful that our predictions could be tested experimentally in the near future and establish BCL as a versatile tool for obtaining tunable topological phases.

\section*{Acknowledgements}

M.G. is supported by the Integrated PhD fellowship of the Indian Institute of Science. A.J. is supported by the INSPIRE fellowship by DST. A.N. acknowledges support from the DST MATRICS grant (MTR/2023/000021).

\bibliographystyle{apsrev4-1}
\bibliography{bibliography}

\end{document}